\documentclass[twocolumn,prl,aps,superscriptaddress,showpacs]{revtex4}
\usepackage{epsfig}
\usepackage{amsmath}
\usepackage{amsthm}
\usepackage{amssymb,braket}
\usepackage{verbatim}
\usepackage{bm}
\usepackage{hyperref}
\usepackage[all]{hypcap}
\usepackage{here}
\usepackage{sidecap}
\usepackage{float}
\usepackage{color}

\begin{document}

\title{Momentum dependent $d_{xz/yz}$ band splitting in LaFeAsO}

\author{S. S. Huh}
\affiliation{Center for Correlated Electron Systems, Institute for Basic Science (IBS), Seoul 08826, Republic of Korea}
\affiliation{Department of Physics and Astronomy, Seoul National University (SNU), Seoul 08826, Republic of Korea}

\author{Y. S. Kim}
\affiliation{Center for Correlated Electron Systems, Institute for Basic Science (IBS), Seoul 08826, Republic of Korea}
\affiliation{Department of Physics and Astronomy, Seoul National University (SNU), Seoul 08826, Republic of Korea}

\author{W. S. Kyung}
\affiliation{Center for Correlated Electron Systems, Institute for Basic Science (IBS), Seoul 08826, Republic of Korea}
\affiliation{Department of Physics and Astronomy, Seoul National University (SNU), Seoul 08826, Republic of Korea}

\author{J. K. Jung}
\affiliation{Center for Correlated Electron Systems, Institute for Basic Science (IBS), Seoul 08826, Republic of Korea}
\affiliation{Department of Physics and Astronomy, Seoul National University (SNU), Seoul 08826, Republic of Korea}

\author{R. Kappenberger}
\affiliation{Leibniz Institute for Solid State and Materials Research, IFW-Dresden, 01069 Dresden, Germany}

\author{S. Aswartham}
\affiliation{Leibniz Institute for Solid State and Materials Research, IFW-Dresden, 01069 Dresden, Germany}

\author{B. B\"{u}chner}
\affiliation{Leibniz Institute for Solid State and Materials Research, IFW-Dresden, 01069 Dresden, Germany}
\affiliation{Institute of Solid State Physics, TU Dresden, 01069 Dresden, Germany}

\author{J. M. Ok}
\affiliation{Center for Artificial Low Dimensional Electronic Systems, Institute of Basic Science, Pohang 790-784, Korea}
\affiliation{Department of Physics, Pohang University of Science and Technology, Pohang 790-784, Korea}

\author{J. S. Kim}
\affiliation{Center for Artificial Low Dimensional Electronic Systems, Institute of Basic Science, Pohang 790-784, Korea}
\affiliation{Department of Physics, Pohang University of Science and Technology, Pohang 790-784, Korea}

\author{C. Dong}
\affiliation{Beijing National Laboratory for Condensed Matter Physics, and Institute of Physics, Chinese Academy of Sciences, Beijing 100190, China}
\affiliation{Collaborative Innovation Center of Quantum Matter, Beijing, China}

\author{J. P. Hu}
\affiliation{Beijing National Laboratory for Condensed Matter Physics, and Institute of Physics, Chinese Academy of Sciences, Beijing 100190, China}
\affiliation{Collaborative Innovation Center of Quantum Matter, Beijing, China}

\author{S. H. Cho}
\affiliation{State Key Laboratory of Functional Materials for Informatics, Shanghai Institute of Microsystem and Information Technology (SIMIT), Chinese Academy of Sciences, Shanghai 200050, People's Republic of China}

\author{D. W. Shen}
\affiliation{State Key Laboratory of Functional Materials for Informatics, Shanghai Institute of Microsystem and Information Technology (SIMIT), Chinese Academy of Sciences, Shanghai 200050, People's Republic of China}

\author{J. D. Denlinger}
\affiliation{Advanced Light Source, Lawrence Berkeley National Laboratory, California 94720, USA}

\author{Y. K. Kim}
\affiliation{Department of Physics, Korea Advanced Institute of Science and Technology, Daejeon 34141, Republic of Korea}
\affiliation{Graduate school of Nanoscience and Technology, Korea Advanced Institute of Science and Technology, Daejeon 34141, Republic of Korea}

\author{C. Kim}
\email{changyoung@snu.ac.kr}
\affiliation{Center for Correlated Electron Systems, Institute for Basic Science (IBS), Seoul 08826, Republic of Korea}
\affiliation{Department of Physics and Astronomy, Seoul National University (SNU), Seoul 08826, Republic of Korea}

\date{\today}

\begin{abstract}
We performed angle-resolved photoemission spectroscopy (ARPES) studies of the electronic structure of the nematic phase in LaFeAsO. Degeneracy breaking between the $d_{xz}$ and $d_{yz}$ hole bands near the $\Gamma$ and $M$ point is observed in the nematic phase. Different temperature dependent band splitting behaviors are observed at the $\Gamma$ and $M$ points. The energy of the band splitting near the $M$ point decreases as the temperature decreases while it has little temperature dependence near the $\Gamma$ point. The nematic nature of the band shift near the $M$ point is confirmed through a detwin experiment using a piezo device. Since a momentum dependent splitting behavior has been observed in other iron based superconductors, our observation confirms that the behavior is a universal one among iron based superconductors.
\pacs{}
\end{abstract}
\maketitle

\section{I. Introduction}
The discovery of iron based superconductors (IBS) over a decade ago brought renewed interest in high T$_{C}$ superconductivity research \cite{intro1,intro2}. In addition to the superconductivity itself, its various phases have attracted much attention due to their possible relation to superconductivity. Among these phases, the nematic phase, a rotational symmetry broken state in the electronic structure, has been intensively studied as it also occurs in other unconventional superconductors  \cite{intronem}. Moreover, the divergent nematic susceptibility at the optimal doping suggests that the nematic fluctuation may play an important role in the formation of the Cooper pairs in IBS \cite{ibsres1,ibsres2}. Therefore, understanding the origin of the nematic phase can be a key to unraveling the mechanism of the unconventional superconductivity in IBS.

A number of angle-resolve photoemission spectroscopy (ARPES) experiments have been conducted to investigate the rotational symmetry broken electronic states. However, interpretations of experimental data differ from each other, making the origin of the nematic phase a more controversial issue. Various scenarios were proposed as the origin of the nematic phase based on ARPES results, such as simple Ferro-orbital order \cite {ming1,ming2}, $d$-wave bond order \cite{dwave}, unidirectional nematic bond order \cite{watson}, and sign reversal order \cite{signreverse}. Among various candidates, the instability of the momentum dependent band splitting which is commonly observed in FeSe, NaFeAs and BaFe$_{2}$As$_{2}$ is favored as the true origin of the nematic phase \cite{huh,pfau,111}.

In resolving such issue, confirming if the behavior of the nematic electronic structure is universal among IBS should be an important step. Despite the intensive research on the nematic phase, there is a lack of APRES study on the nematic phase of LaFeAsO even though it is the first IBS that was discovered \cite{feng,peng}. Therefore, investigating electronic structure of the LaFeAsO nematic phase is an important step towards finding the origin of the nematic phase. In this work, we performed temperature dependent ARPES experiments on twinned and detwinned LaFeAsO crystals. The results show a nematic behavior similar to other IBS: finite near the $M$ point but very small near the $\Gamma$ point \cite{huh,pfau,111}. Our observation of the momentum dependent nematic band splitting establishes a universal nematic behavior in IBS.

\section{II. Experiment}

High quality LaFeAsO single crystals were synthesized by using solid state crystal growth technique \cite{growth}. The structural and magnetic transition temperatures ($T_S$ and $T_N$) were found to be around 145 and 127 K, respectively \cite{growth, character}. ARPES measurements were performed at the beam line 4.0.3 of the Advanced Light source (ALS) and beam line 03U of the Shanghai Synchrotron Radiation Facility (SSRF), and also with a lab-based system at Seoul National University (SNU). Piezo detwin ARPES experiments were performed with the system at SNU. 150 V bias voltage was applied to the piezo device to detwin samples. All spectra were acquired with VG-Scienta electron analyzers. The samples were cleaved in an ultrahigh vacuum better than $5 \times 10^{-11}$ torr. To minimize the aging effect, all the data were taken within 8 hours after cleave.

\section{III. Results and discussion}
\begin{figure}[h]
\centering \epsfxsize=8.6cm \epsfbox{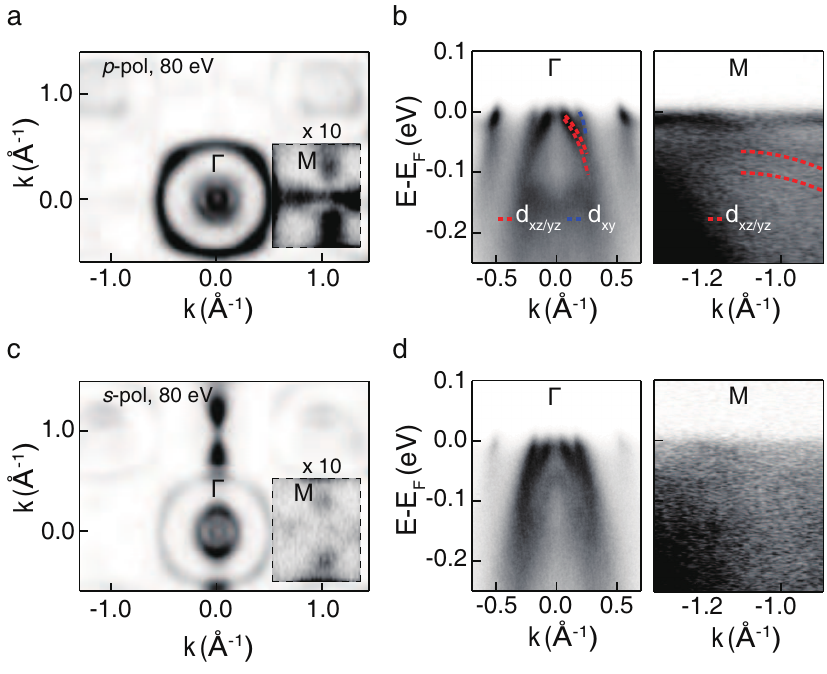} \caption{(Color online). (a) Fermi surface map taken with 80 eV and $p$-polarized light. The inset shows the intensity near the $M$ point multiplied by 10 to show the details. (b) Corresponding band structure at the $\Gamma$ and $M$ points along the $\Gamma$-$M$ direction. The red and blue dashed line indicate dispersions of $d_{xz/yz}$ and $d_{xy}$ bands, respectively. (c) and (d) Similar measurements but with $s$-polarized light. All the data were taken at 30 K.}
\label{fig1}
\end{figure}

Figs. 1(a) and 1(c) show Fermi surface maps of LaFeAsO taken with $p$- and $s$-polarized 80 eV light, respectively. Considering all the features of the two Fermi surface maps taken with different polarizations, we determine that the Fermi surface consists of three circular pockets around the $\Gamma$ point and two peanut like pockets around the $M$ point. Due to the low intensity near the $M$ point, the intensity of the Fermi surface near the $M$ point is multiplied by 10 and depicted in the inset. The observed Fermi surface topology is consistent with that from previous ARPES studies on LaFeAsO \cite{feng,peng,sm,dong}. Band structures near $\Gamma$ and $M$ are shown in Figs. 1(b) and 1(d) for $p$- and $s$-polarizations, respectively. Shown in Fig. 1(b) are the band dispersions as well as their orbital characters, determined based on tight binding calculation (See Supplementary). It is noteworthy that the bands which cannot be identified with the calculation results are surface states.

\begin{figure*}
\centering \epsfxsize=17.2cm \epsfbox{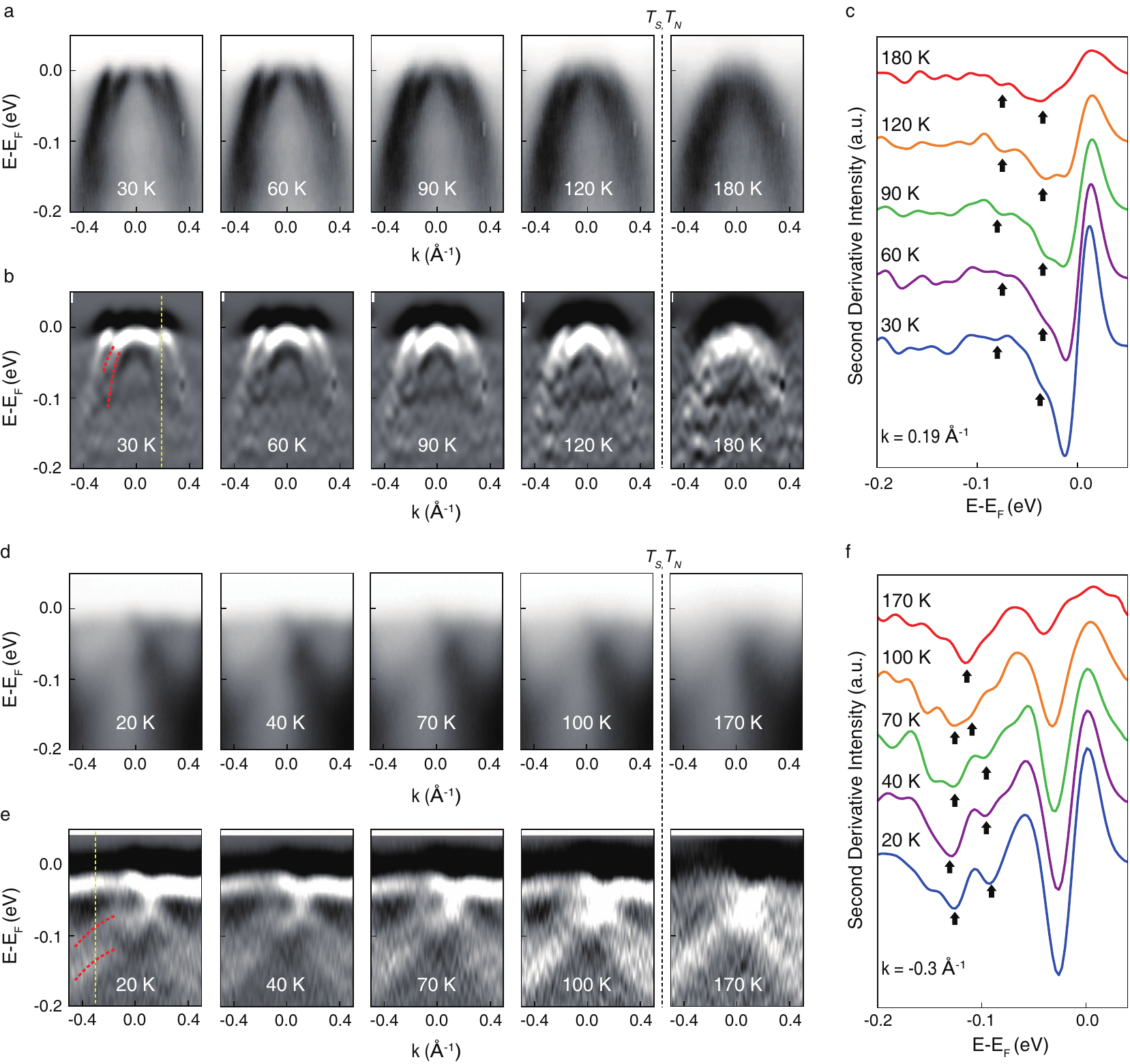} \caption{(Color online). (a) Temperature dependent ARPES data and (b) its second derivative near the $\Gamma$ point taken with 80 eV photon. The red dashed lines indicate dispersions of $d_{xz/yz}$ bands. (c) Temperature dependent energy distribution curves (EDCs) at $k=0.19{\AA}^{-1}$, indicated by the yellow dashed line in Fig. 2(b). Peak positions of $d_{xz}$ and $d_{yz}$ are indicated by black arrows. (d) Temperature dependent ARPES data and (e) its second derivative data near the $M$ point taken with 21.2 eV photon energy, for which the cross section is higher. (f) Temperature dependent EDCs at $k=-0.3{\AA}^{-1}$, indicated by the yellow dashed line in Fig. 2(e). }
\label{fig2}
\end{figure*}

We performed temperature dependent experiments presented to gain more information. Experiments were done 1 hour after cleaving the sample at 30 K to remove the surface reconstruction effect \cite{peng}. The temperature evolution of $d_{xz}$/$d_{yz}$ splitting near the $\Gamma$ point can be seen in the high symmetry cut and its second derivative data shown in Figs. 2(a) and 2(b), respectively. Temperature dependent high symmetry cut data shows that the band dispersions around the $\Gamma$ point hardly change except the thermal broadening. That is, $d_{xz/yz}$ hole band splitting near $\Gamma$, as indicated by the red dashed lines, appears to remain almost unchanged over the temperature range we studied. The energy splitting size may be obtained from the temperature dependent energy distribution curves (EDCs) of the second derivative data in Fig. 2(c). The EDCs are from $k=0.19{\AA}^{-1}$ as indicated by the yellow dotted line in Fig. 1(b). Peak positions of $d_{xz}$ and $d_{yz}$ hole bands are indicated by the arrows. The band splitting at 30 K is about 50 meV and changes very little with the temperature, remaining finite even above $T_S$. That is, the $d_{xz}$/$d_{yz}$ hole band splitting near the $\Gamma$ point is not sensitive to the nematic phase transition.

In previous studies on other IBS, the $d_{xz}$/$d_{yz}$ hole band splitting near the $M$ point has been intensively studied because of its large temperature dependence \cite{ming1,ming2,dwave,signreverse,huh,peng,pfau}. For that reason, we also focus on the temperature dependence of the electronic structure near the $M$ point. Measurements were done with 21.2 eV photon energy where the cross section is higher for $M$. Temperature dependent data and their second derivatives are plotted in Figs. 2(d) and 2(e). Split $d_{xz}$ and $d_{yz}$ hole bands are observed at low temperatures as indicated by red dashed lines. In contrast to the temperature independent behavior near the $\Gamma$ point, the splitting between the two hole bands gradually decreases as the temperature increases. The two bands finally merge each other to a single band above $T_S$. The temperature dependence of band splitting may be observed more clearly in the EDCs of second derivative data at $k=-0.3{\AA}^{-1}$ in Fig. 2(f). The splitting energy at a low temperature is about 40 meV and vanishes above the nematic phase transition temperature $T_S$.

Even though the temperature dependent behavior strongly indicates that the band splitting near $M$ is due to the nematic order, a true confirmation of the nematic nature should come from observation of dependent shift. In previous ARPES studies on detwinned IBS \cite{ming1,ming2,huh}, nematic band shift has been observed in which the $d_{yz}$ ($d_{xz}$) hole band from a single domain shifts upward (downward) along $\Gamma$-$M_{X}$ ($M_{Y}$). Such band shifts lead to a simple energy splitting between the $d_{xz}$ and $d_{yz}$ hole bands in the twinned sample due to the mixed signals from two perpendicularly arranged domains in the nematic phase. It is reasonable to speculate that observed band splitting near the $M$ point in our data is also from the superposition of the $d_{yz}$ hole band along $\Gamma$-$M_{X}$ and the $d_{xz}$ hole band along $\Gamma$-$M_{Y}$ in LaFeAsO. This can be clarified through detwin experiments.

\begin{figure}
\centering \epsfxsize=8.6cm \epsfbox{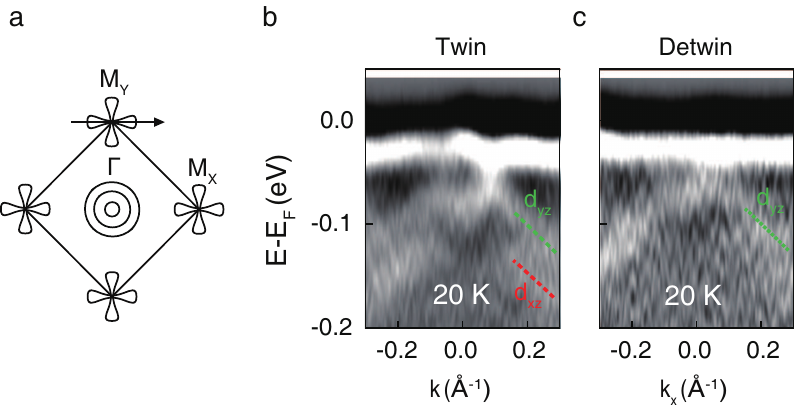} \caption{(Color online). (a) Schematic illustration of experimental geometry. High symmetry cut of the (b) twinned and (c) detwinned sample along $k_{x}$-direction near the $M_{Y}$ point. The red and green dashed line indicated dispersion of $d_{xz}$ and $d_{yz}$ band, respectively. Tensile strain is transmitted to sample by applying 150 V to the piezo strain device.}
\label{fig3}
\end{figure}

Detwinning was done by using a piezo stack based strain device which was adopted in previous FeSe studies \cite{huh}. Tensile strain is transmitted to the sample glued to the piezo device when voltage is applied to the device. In our experiment, we applied 150 V to the piezo device at 200 K and cooled down the sample. Band dispersion near the $M_{Y}$ point along the $k_{x}$-direction in twinned and detwinned samples are presented in Figs. 3(b) and (c), respectively. In the detwin data, only the upper hole band ($d_{yz}$) without the lower hole band ($d_{xz}$) is seen in comparison to the twin data. Note that the band dispersion in this experimental geometry is the same as the one obtained along the $k_{x}$-direction at $M_{X}$ due to the band folding; $M_Y$ was chosen because the intensity is higher. Therefore, considering the normal state band dispersions as well as our detwin experiment result, we conclude that LaFeAsO also has nematic band shifts similar to other IBS \cite{nor}.

The discovery of the nematic band shift near $M$ was taken to be an evidence for a Ferro-orbital order which in turn was considered to be the origin of the nematic phase. It is theoretically suggested that a nearly constant band splitting is expected in the entire Brillouin zone when Ferro-orbital order exists \cite{ccl,ccc}. However, ARPES results show that the band splitting has momentum dependence for all IBS systems including LaFeAsO, which excludes the Ferro-orbital scenario as the origin. Rather than the Ferro-orbital order, the instability for the observed universal momentum dependent band splitting should be the true origin of the nematic phase. There are proposals such as different form of orbital degree of freedom \cite{hopping}, Pomeranchuk instability \cite{pom} and spin degree of freedom \cite{orbdepspin,spin1,spin2} that require momentum dependence. Finding out which one of these is truly responsible for the nematic phase requires full understanding of the nematic electronic structure. As proposed very recently \cite{huh,ming3}, in addition to consideration of $d_{xz}$ and $d_{yz}$ hole bands, evolution of $d_{xz}$/$d_{yz}$ electron bands, spin orbit coupling and role of $d_{xy}$ band should be elucidated to resolve the issue.

\section{IV. CONCLUSION}
In conclusion, we report the detailed electronic structure of the nematic phase in LaFeAsO. We observe momentum dependent $d_{xz}$/$d_{yz}$ band splitting behavior near the $\Gamma$ and $M$ points in temperature dependent experiments. The splitting size near the $M$ point decreases as the temperature increases, while it is less temperature sensitive near the $\Gamma$ point. We confirmed the nematic nature of the band splitting near $M$ by piezo detwin method. Our observation establishes the momentum dependent splitting as the universal behavior of the electronic structure in the nematic phase of IBS. Our result suggests that instability causes the observed universal momentum dependent band splitting should be the true origin of the nematic phase.

\section{Acknowledgment}
Authors would like to thank H. Pfau and Y. Ishida for helpful discussions. This work is supported by IBS-R009-G2 through the IBS Center for Correlated Electron Systems. Theoretical part was supported by grants from CAS (XDB07000000). Part of this research used Beamline 03U of the Shanghai Synchron Radiation Facility, which is supported by ME2 project under contract No. 11227902 from National Natural Science Foundation of China. The work at Pohang University of Science and Technology (POSTECH) was supported by IBS (no. IBS-R014-D1) and the NRF through the SRC (No. 2018R1A5A6075964) and the Max Planck-POSTECH Center (No. 2016K1A4A4A01922028). The work at IFW was supported by the (DFG)  through  the Priority Program SPP 1458. The work at Korea Advanced Institute of Science and Technology (KAIST) was supported by NRF through National R$\&$D Program (No. 2018K1A3A7A09056310), Creative Materials Discovery Program (No. 2015M3D1A1070672), and Basic Science Resource Program (No. 2017R1A4A1015426, No. 2018R1D1A1B07050869). The Advanced Light Source is supported by the Office of Basic Energy Sciences of the US DOE under Contract No. DE-AC02-05CH11231.

\end{document}